\documentclass[aps,pra,superscriptaddress,floatfix,showpacs,10pt,twocolumn]{revtex4}
\usepackage[dvips]{graphicx}
\usepackage{amsmath}
\begin{document}
\title[Quantum Information Processing Without Joint Measurement]{Quantum Information Processing Without Joint Measurement}
\author{Ming Yang}
\email{mingyang@ahu.edu.cn}
\author{Zhuo-Liang Cao}
\email{zlcao@ahu.edu.cn} \affiliation{School of Physics {\&}
Material Science, Anhui University, Hefei, 230039, People's
Republic of China} 

\begin{abstract}
We propose a linear optical scheme for the teleportation of
unknown ionic states, the entanglement concentration for
nonmaximally entangled states for ions via entanglement swapping
and the remote preparation for ionic entangled states. The unique
advantage of the scheme is that the joint Bell-state measurement
needed in the previous schemes is not needed in the current
scheme, i.e. the joint Bell-state measurement has been converted
into the product of separate measurements on single ions and
photons. In addition, the current scheme can realize the quantum
information processes for ions by using linear optical elements,
which simplify the implementation of quantum information
processing for ions.
\end{abstract}

\maketitle

\section{INTRODUCTION}

Joint measurements play an important role in quantum information
processing(QIP). In quantum teleportation, an unknown quantum
state will be sent from sender to receiver via a quantum channel
with the help of classical communication. During this process, the
sender will operate a joint Bell-state measurement on the
unknown-state particle and one of the entangled particles she
possesses~\cite{bennett}. In entanglement swapping, there are
usually three spatially separate users, and two of them have
shared one pair of entangled particles with the third user. Then
the third user will operate a joint Bell-state measurement on the
two particles he possesses. Corresponding to the measurement
result, the two particles possessed by the two spatially separate
users will collapse into an entangled state without any
entanglement before the joint measurement~\cite{swapp}. In remote
state preparation~\cite{rsp1}, if we want to prepare some
entangled state remotely via two pairs of entangled particles as
the quantum channel, the joint measurement is also a
necessity~\cite{rsp2}. Like the above-mentioned entanglement
swapping, there are usually three spatially separate users, and
two of them have shared one pair of entangled particles with the
third user. Then the third user user will operate a special type
of joint measurement, which corresponds to the entangled state the
third user want to prepare remotely, on the two particles he
possesses. After the third user informs the first or the second
user the measurement results, the particles of the first and the
second user are prepared in the entangled state that the third
user want to prepare remotely. In addition, joint measurements are
also needed in the quantum dense coding~\cite{densecoding}. The
sender and the receiver share an entangled pairs, and the sender
will encode the 2-bit classical information on the entangled
particle he possesses by operating four possible single-bit
operations on the particle. Then this particle(one qubit) will be
sent to the receiver. The receiver will operate a joint Bell-state
measurement on the two particles to decode the 2-bit classical
information the sender encoded.

Form the above analysis, joint measurements are needed in quantum
teleportation, entanglement swapping, remote preparation of
entangled state via two pairs of entangled particles as the
quantum channel and quantum dense coding, etc. But it is very
difficult to realize joint measurements in experiment. Usually,
joint measurements will be converted into the product of separate
measurements on single particle~\cite{pan, zheng, zhaozhi, ye}.
For the photon case, teleportation of unknown polarization state
of photons has been realized in experiment~\cite{Bouwmeester},
where the joint Bell-state measurement has been converted into the
product of separate measurements on single photon by using the
linear optical elements, such as polarization beam splitters and
photon detectors. But in this experimental scheme, the four Bell
states can not be discriminated conclusively and completely. So
Zhao \emph{et al} proposed an alternative scheme for the
conclusive discrimination of the four Bell states of photons with
the help of the ancillary entangled pairs of
photons~\cite{zhaozhi}. In cavity QED domain, the Bell-state
measurement on atoms has been converted into the separate
measurements on single atom by using the controlled-NOT (C-NOT)
gate operations~\cite{zheng}, where the dispersive interaction
between two atoms and a cavity mode plays an important role. To
avoid the difficulty of the C-NOT gate operations, Zheng and Ye
all proposed the schemes to teleport an unknown atomic state
without Bell-state measurement, and the interaction between atoms
and cavity modes decomposes the Bell states into product states.

From the experimental point of view, Bell-state measurements have
been realized for the photon case. However, because it is
difficult to realize Bell-state measurements for atomic (ionic)
states in experiment~\cite{experimentele1, experimentele2}, the
implementation of quantum teleportation~\cite{experimentele1,
experimentele2}, entanglement swapping, remote preparation of
entangled state and quantum dense coding for atomic (ionic) states
are all not easy. In our previous contribution, we have proposed
the entanglement swapping scheme for atomic system without joint
measurement~\cite{me1, me2}, where the interaction (resonant and
nonresnant cases) between atoms and cavity modes replaces the
Bell-state analyzer. But, due to the complexity of the cavity QED
techniques, it is difficult to realize these schemes in
experiment. So, in this paper, we will propose an alternative
scheme for some quantum information processes for ions, such as,
quantum teleportation, entanglement concentration via entanglement
swapping and remote preparation of entangled states, etc. We will
use the linear optical elements. The main setup of the scheme is a
Mach-Zehnder interferometers (MZI) with two ions placed on the two
arms of the MZI. We can decide whether the scheme succeed or not
by operating single photon measurements on the two output ports of
the MZI. The unique advantage of the current scheme is that the
quantum information processing for ions can be realized by using
linear optical elements, which decreases the difficulty of the
experimental implementation. In addition, by using the MZI, the
joint measurement are all decomposed into single photon
measurements and single ion measurements, so the current scheme
avoids the difficulty of realizing joint measurements.

\section{QUANTUM TELEPORTATION OF UNKNOWN IONIC STATES VIA LINEAR OPTICS}

\begin{figure}
\includegraphics[scale=0.27, angle=270]{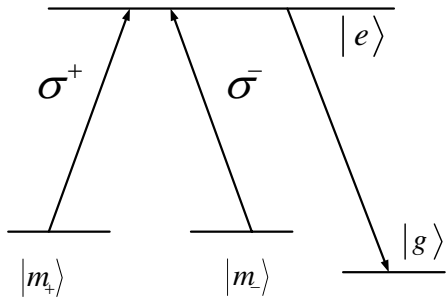}
\caption{\label{level}Level configuration of the ions used in the
scheme. The ions, which are in the degenerate states
$|m_{+}\rangle\textrm{ and }|m_{-}\rangle$, can be excited into
the unstable excited state $|e\rangle$ by absorbing one
$\sigma^{+}\textrm{ or }\sigma^{-}$ polarized photon, then it can
decay to the stable ground state $|g\rangle$ with a scattered
photon rapidly.}
\end{figure}

Consider the three-level ionic system, where
$|m_{+}\rangle\textrm{ and }|m_{-}\rangle$ are two degenerate
metastable states of ions, and $|e\rangle$ is the excited state.
The level configuration of the ions is depicted in
Fig.\ref{level}.
The ions can be excited from $|m_{+}\rangle\textrm{ or
}|m_{-}\rangle$ to the excited states $|e\rangle$ by absorbing one
$\sigma^{+}\textrm{ or }\sigma^{-}$ circular polarization photon
with unit efficiency. Because the excited state $|e\rangle$ is not
a stable one, the ions in that state will decay rapidly to the
stable ground state $|g\rangle$ with a scattered photon
$|S\rangle$. This process can be expressed as~\cite{me3, nqi}:
\begin{equation}\label{scattering}
a_{\pm}^{+}|0\rangle|m_{\pm}\rangle\longrightarrow|S\rangle|g\rangle.
\end{equation}

Suppose the unknown state of ion 1 to be teleported is:
\begin{equation}\label{statetobetele}
|\Psi\rangle_{1}=\alpha|m_{+}\rangle_{1}+\beta|m_{-}\rangle_{1},
\end{equation}
where $\alpha, \beta$ are normalization coefficients, and
$|\alpha|^{2}+|\beta|^{2}=1$. Without loss of generality, we can
assume that $\alpha, \beta$ are all real numbers.

\begin{figure}
\includegraphics[scale=0.67]{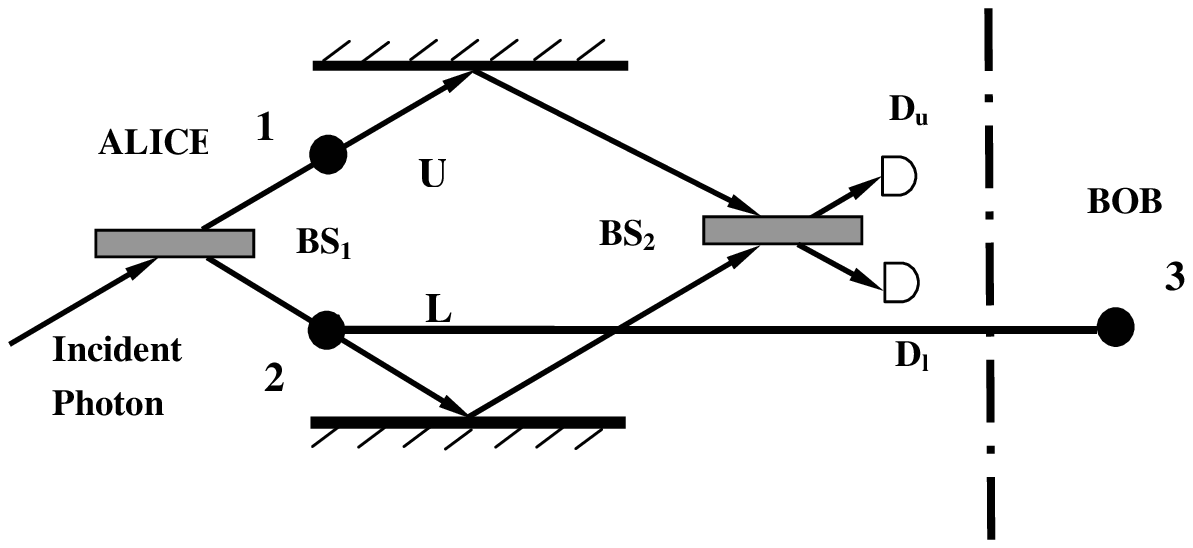}
\caption{\label{setupfortele} The setup for teleportation scheme.
Alice places ions 1, 2 on the upper arm and lower arm of the MZI,
respectively. One $\sigma^{+}$ polarized photon will be
superimposed on the first beam splitter (BS$_{1}$). After the
BS$_{1}$ the photon will take two possible pathes (\emph{u
}denotes the upper path and \emph{l} denotes the lower one).
Reflected by two mirrors, the two possible pathes will be
recombined at the second beam splitter (BS$_{2}$). Single photon
detection will be operated on the two output ports of the MZI
after the BS$_{2}$ . If the right lower output port detector
($D_{l}$) fires, the teleportation can succeed.}
\end{figure}

Before teleportation, the sender (Alice) and the receiver (Bob)
share a maximally entangled pair of ions 2, 3:
\begin{equation}\label{channelfortele}
|\Psi\rangle_{23}=\frac{1}{\sqrt{2}}(|m_{+}\rangle_{2}|m_{+}\rangle_{3}+|m_{-}\rangle_{2}|m_{-}\rangle_{3}).
\end{equation}
Alice possesses the ions 1, 2, and Bob has access to ion 3. To
complete the teleportation of unknown state of ion 1, a MZI will
be introduced at Alice's location. The ions 1, 2 will be placed on
upper arm and lower arm of the MZI, respectively, by using
trapping technology~\cite{trapping}. One $\sigma^{+}$ polarized
photon will be superimposed on the MZI from the left lower input
port. The main setup is depicted in Fig.\ref{setupfortele}.
The effect of the BS on the input photon can be expressed as:
\begin{subequations}\label{effectofbs}
\begin{equation}
a_{l,\pm}^{+}|0\rangle\stackrel{\textrm{BS}}{\longrightarrow}
\frac{1}{\sqrt{2}}(a_{u,\pm}^{+}+ia_{l,\pm}^{+})|0\rangle,
\end{equation}
\begin{equation}
a_{u,\pm}^{+}|0\rangle\stackrel{\textrm{BS}}{\longrightarrow}
\frac{1}{\sqrt{2}}(a_{l,\pm}^{+}+ia_{u,\pm}^{+})|0\rangle,
\end{equation}
\end{subequations}
where $l\textrm{ and }u$ denote the lower optical path and the
upper optical path, respectively. $a_{l,\pm}^{+}|0\rangle\textrm{
and } a_{u,\pm}^{+}|0\rangle$ denote two photons, and $\pm$
denotes the direction of polarization. From Eq.(\ref{effectofbs})
we found that there will be a $\frac{\pi}{2}$ phase shift between
the input photon and the reflected wave function, and the
transparent part is synchronized with the input photon. The BS
takes no effect on the polarization of the input photon. Because
the ions 1, 2 have been placed on the two optical pathes (\emph{u}
and \emph{l}, respectively), the ions will interact with the
photon. The interaction will generate a shift of interference
after the BS$_{2}$, then Alice will detect the two output ports of
the MZI to check whether the teleportation succeeds or not. These
are critical to the teleportation and the other quantum
information processes to be presented in the current scheme.

Before teleportation the total state of the system is:
\begin{eqnarray}\label{totalfortele}
|\Psi\rangle_{total}&=&\frac{1}{\sqrt{2}}a_{l,+}^{+}|0\rangle(\alpha|m_{+}\rangle_{1}|m_{+}\rangle_{2}|m_{+}\rangle_{3}\nonumber\\
&+&\alpha|m_{+}\rangle_{1}|m_{-}\rangle_{2}|m_{-}\rangle_{3}\nonumber\\
&+&\beta|m_{-}\rangle_{1}|m_{+}\rangle_{2}|m_{+}\rangle_{3}\nonumber\\
&+&\beta|m_{-}\rangle_{1}|m_{-}\rangle_{2}|m_{-}\rangle_{3}).
\end{eqnarray}

To analyze the evolution of the total system, we will consider the
evolution of the following four product states of ions 1, 2:
\begin{subequations}\label{basicevolution}
\begin{eqnarray}
&a_{l,+}^{+}|0\rangle|m_{+}\rangle_{1}|m_{+}\rangle_{2}
\xrightarrow{BS_{1}, Ions 1, 2, BS_{2}}\nonumber\\
&\frac{1}{\sqrt{2}}
(|S\rangle_{1}|g\rangle_{1}|m_{+}\rangle_{2}+i|m_{+}\rangle_{1}|S\rangle_{2}|g\rangle_{2}),
\end{eqnarray}
\begin{eqnarray}
&a_{l,+}^{+}|0\rangle|m_{+}\rangle_{1}|m_{-}\rangle_{2}
\xrightarrow{BS_{1}, Ions 1, 2, BS_{2}}\nonumber\\
&\frac{1}{\sqrt{2}}|S\rangle_{1}|g\rangle_{1}|m_{-}\rangle_{2}\nonumber\\
&+\frac{i}{2}(a_{u,+}^{+}+ia_{l,+}^{+})|0\rangle|m_{+}\rangle_{1}|m_{-}\rangle_{2}
\end{eqnarray}
\begin{eqnarray}
&a_{l,+}^{+}|0\rangle|m_{-}\rangle_{1}|m_{+}\rangle_{2}
\xrightarrow{BS_{1}, Ions 1, 2, BS_{2}}\nonumber\\
&\frac{i}{\sqrt{2}}|m_{-}\rangle_{1}|S\rangle_{2}|g\rangle_{2}\nonumber\\
&+\frac{1}{2}(a_{l,+}^{+}+ia_{u,+}^{+})|0\rangle|m_{-}\rangle_{1}|m_{+}\rangle_{2},
\end{eqnarray}
\begin{eqnarray}
&a_{l,+}^{+}|0\rangle|m_{-}\rangle_{1}|m_{-}\rangle_{2}
\xrightarrow{BS_{1}, Ions 1, 2, BS_{2}}\nonumber\\
&ia_{u,+}^{+}|0\rangle|m_{-}\rangle_{1}|m_{-}\rangle_{2}.
\end{eqnarray}
\end{subequations}
So, after the operation of MZI, the state of the total system will
evolve into:
\begin{eqnarray}\label{afterevolutionfortele}
|\Psi\rangle^{'}_{total}&=&\frac{1}{\sqrt{2}}\{\alpha\frac{1}{\sqrt{2}}(|S\rangle_{1}|g\rangle_{1}|m_{+}\rangle_{2}\nonumber\\
&+&i|m_{+}\rangle_{1}|S\rangle_{2}|g\rangle_{2})|m_{+}\rangle_{3}+\alpha[\frac{1}{\sqrt{2}}|S\rangle_{1}|g\rangle_{1}|m_{-}\rangle_{2}\nonumber\\
&+&\frac{i}{2}(a_{u,+}^{+}+ia_{l,+}^{+})|0\rangle|m_{+}\rangle_{1}|m_{-}\rangle_{2}]|m_{-}\rangle_{3}\nonumber\\
&+&\beta[\frac{i}{\sqrt{2}}|m_{-}\rangle_{1}|S\rangle_{2}|g\rangle_{2}\nonumber\\
&+&\frac{1}{2}(a_{l,+}^{+}+ia_{u,+}^{+}) |0\rangle|m_{-}\rangle_{1}|m_{+}\rangle_{2}]|m_{+}\rangle_{3}\nonumber\\
&+&i{\beta}a_{u,+}^{+}|0\rangle|m_{-}\rangle_{1}|m_{-}\rangle_{2}|m_{-}\rangle_{3}\}.
\end{eqnarray}
If the photon detector at the right lower output port $D_{l}$
fires, the state of the system collapse into:
\begin{eqnarray}\label{teleportedstate}
|\Psi\rangle_{123}&=&\frac{1}{2\sqrt{2}}(-\alpha|m_{+}\rangle_{1}|m_{-}\rangle_{2}|m_{-}\rangle_{3}\nonumber\\
&+&\beta|m_{-}\rangle_{1}|m_{+}\rangle_{2}|m_{+}\rangle_{3}).
\end{eqnarray}
Then Alice will measure the ions 1, 2 in the basis:
\begin{subequations}\label{newbasis}
\begin{equation}
|+\rangle=\frac{1}{\sqrt{2}}(|m_{+}\rangle+|m_{-}\rangle),
\end{equation}
\begin{equation}
|-\rangle=\frac{1}{\sqrt{2}}(|m_{+}\rangle-|m_{-}\rangle).
\end{equation}
\end{subequations}
For the results $|+\rangle_{1}|+\rangle_{2}$,
$|-\rangle_{1}|-\rangle_{2}$, the ion 3 will be left in the state
$-\alpha|m_{-}\rangle_{3}+\beta|m_{+}\rangle_{3}$, so a
$\sigma_{y}$ operation is needed to transfer the state of ion 3 to
the state of ion 1. For the results $|+\rangle_{1}|-\rangle_{2}$,
$|-\rangle_{1}|+\rangle_{2}$, the ion 3 will be left in the state
$\alpha|m_{-}\rangle_{3}+\beta|m_{+}\rangle_{3}$, so a
$\sigma_{x}$ operation is needed to transfer the state of ion 3 to
the state of ion 1. The total success probability of the
teleportation scheme is $1/8$. Although the successful probability
is smaller than 1.0, the current scheme does not need the joint
Bell-state measurement and the complex ion trap techniques, which
will simplify the implementation of the scheme. The current scheme
for the teleporation of one qubit unknown state can be generalized
to the multi-qubit unknown state case in a straight forward way.
This will be discussed in more detail in other papers.

\section{ENTANGLEMENT CONCENTRATION FOR IONIC ENTANGLED STATE VIA ENTANGLEMENT SWAPPING}

Suppose there are three spatially separate users Alice, Bob and
Cliff. Alice and Bob all share a pair of nonmaximally entangled
ions with Cliff. Alice has access to ion 1, Bob has access to ion
4, and ions 2 and 3 are all at Cliff's location. The entangled
state of ions 1, 2 and the entangled state of ions 3, 4 are:
\begin{subequations}\label{channelforswapping}
\begin{equation}
|\Phi\rangle_{12}=\alpha|m_{+}\rangle_{1}|m_{+}\rangle_{2}+\beta|m_{-}\rangle_{1}|m_{-}\rangle_{2},
\end{equation}
\begin{equation}
|\Phi\rangle_{34}=a|m_{+}\rangle_{3}|m_{+}\rangle_{4}+b|m_{-}\rangle_{3}|m_{-}\rangle_{4}.
\end{equation}
\end{subequations}
Where $\alpha, \beta, a, b$ are normalization coefficients, and
$|\alpha|^{2}+|\beta|^{2}=1$, $|a|^{2}+|b|^{2}=1$. Without loss of
generality, we can assume that $\alpha, \beta, a, b$ are all real
numbers.

Before swapping, the state of the total system is:
\begin{eqnarray}\label{totalforswapping}
|\Phi\rangle_{total}&=&a_{l,+}^{+}|0\rangle({\alpha}a|m_{+}\rangle_{1}|m_{+}\rangle_{2}|m_{+}\rangle_{3}|m_{+}\rangle_{4}\nonumber\\
&+&{\alpha}b|m_{+}\rangle_{1}|m_{+}\rangle_{2}|m_{-}\rangle_{3}|m_{-}\rangle_{4}\nonumber\\
&+&{\beta}a|m_{-}\rangle_{1}|m_{-}\rangle_{2}|m_{+}\rangle_{3}|m_{+}\rangle_{4}\nonumber\\
&+&{\beta}b|m_{-}\rangle_{1}|m_{-}\rangle_{2}|m_{-}\rangle_{3}|m_{-}\rangle_{4}).
\end{eqnarray}
To construct entanglement between ion 1 and ion 4, Cliff will
introduce the MZI as in section II, and put the ions 2, 3 on the
upper and lower arm of the MZI, respectively. Then One
$\sigma^{+}$ polarized photon will be superimposed on the MZI from
the left lower input port. From the evolution in
Eq.(\ref{basicevolution}), we can give the evolution caused by the
MZI:
\begin{eqnarray}\label{afterevolutionforswapping}
&&|\Psi\rangle^{'}_{total}={\alpha}a\frac{1}{\sqrt{2}}(|S\rangle_{2}|g\rangle_{2}|m_{+}\rangle_{3}\nonumber\\
&+&i|m_{+}\rangle_{2}|S\rangle_{3}|g\rangle_{3})|m_{+}\rangle_{1}|m_{+}\rangle_{4}+{\alpha}b[\frac{1}{\sqrt{2}}|S\rangle_{2}|g\rangle_{2}|m_{-}\rangle_{3}\nonumber\\
&+&\frac{i}{2}(a_{u,+}^{+}+ia_{l,+}^{+})|0\rangle|m_{+}\rangle_{2}|m_{-}\rangle_{3}]|m_{+}\rangle_{1}|m_{-}\rangle_{4}\nonumber\\
&+&{\beta}a[\frac{i}{\sqrt{2}}|m_{-}\rangle_{2}|S\rangle_{3}|g\rangle_{3}\nonumber\\
&+&\frac{1}{2}(a_{l,+}^{+}+ia_{u,+}^{+}) |0\rangle|m_{-}\rangle_{2}|m_{+}\rangle_{3}]|m_{-}\rangle_{1}|m_{+}\rangle_{4}\nonumber\\
&+&i{\beta}ba_{u,+}^{+}|0\rangle|m_{-}\rangle_{2}|m_{-}\rangle_{3}|m_{-}\rangle_{1}|m_{-}\rangle_{4}.
\end{eqnarray}
Then Cliff will detect the two output ports of the MZI. If the
photon detector at the right lower output port $D_{l}$ fires, the
state of the total system will collapse into:
\begin{eqnarray}\label{totalforswappingafterevolution}
|\Phi\rangle_{1234}&=&\frac{1}{2}(-{\alpha}b|m_{+}\rangle_{1}|m_{+}\rangle_{2}|m_{-}\rangle_{3}|m_{-}\rangle_{4}\nonumber\\
&+&{\beta}a|m_{-}\rangle_{1}|m_{-}\rangle_{2}|m_{+}\rangle_{3}|m_{+}\rangle_{4}).
\end{eqnarray}
If we let $\alpha=a$ $\beta=b$, the state in
Eq.(\ref{totalforswappingafterevolution}) becomes:
\begin{eqnarray}\label{swappedstate}
|\Phi\rangle_{1234}&=&\frac{1}{2}ab(|m_{-}\rangle_{1}|m_{-}\rangle_{2}|m_{+}\rangle_{3}|m_{+}\rangle_{4}\nonumber\\
&-&|m_{+}\rangle_{1}|m_{+}\rangle_{2}|m_{-}\rangle_{3}|m_{-}\rangle_{4}),
\end{eqnarray}
which is a four-ion maximally entangled states. If Cliff measures
the ions 2, 3 in the $|\pm\rangle$ basis as in section II, the
ions 1, 4 will be left in maximally entangled state
$\frac{1}{\sqrt{2}}(|m_{-}\rangle_{1}|m_{+}\rangle_{4}-|m_{+}\rangle_{1}|m_{-}\rangle_{4})$
for the results $|+\rangle_{2}|+\rangle_{3}$,
$|-\rangle_{2}|-\rangle_{3}$. If the results are
$|+\rangle_{2}|-\rangle_{3}$, $|-\rangle_{2}|+\rangle_{3}$, the
ions 1, 4 will be left in the maximally entangled state
$\frac{1}{\sqrt{2}}(|m_{-}\rangle_{1}|m_{+}\rangle_{4}+|m_{+}\rangle_{1}|m_{-}\rangle_{4})$.
The total success probability is $\frac{1}{2}|a|^{2}(1-|a|^{2})$.

So the two ions 1, 4, which have never interacted before, are left
in an entangled state via entanglement swapping. Furthermore, the
initial nonmaximally entangled states in
Eq.(\ref{channelforswapping}) have been concentrated into a
maximally entangled state via entanglement swapping~\cite{bose}.
In our previous contributions~\cite{me1, me2}, we also realized
the entanglement concentration via entanglement swapping in cavity
QED, and the success probabilities of them are bigger than that of
the current scheme. But, the current scheme uses the linear
optical elements, which can be realized within the current
technology easily. In addition, four-ion maximally entangled
states can be generated in the current scheme. In one of our
previous contributions, we used the similar setup, and realized
the purification and concentration of nonmaximally entangled ionic
states~\cite{me3}. But two MZIs have been used there. In our
current scheme, the concentration can be realized by using the MZI
only once. So the current scheme is more efficient than the
previous one.

\section{REMOTE PREPARATION OF ENTANGLED STATES}

In section III, if Cliff want to prepare an entangled state:
\begin{equation}\label{stateofrsp}
|\Phi\rangle_{14}=m|m_{+}\rangle_{1}|m_{-}\rangle_{4}+n|m_{-}\rangle_{1}|m_{+}\rangle_{4},
\end{equation}
on ions 1, 4 remotely, Cliff will measure the state of ions 2, 3
in Eq. (\ref{swappedstate}) in the basis $|\pm^{'}\rangle$:
\begin{subequations}\label{newbasisforrsp}
\begin{equation}
|+^{'}\rangle=\nu|m_{+}\rangle+\mu|m_{-}\rangle,
\end{equation}
\begin{equation}
|-^{'}\rangle=-\mu|m_{+}\rangle+\nu|m_{-}\rangle.
\end{equation}
\end{subequations}
Where $m=\frac{\nu^{2}}{\sqrt{\nu^{4}+\mu^{4}}}$,
$n=\frac{\mu^{2}}{\sqrt{\nu^{4}+\mu^{4}}}$. After the measurement,
the ions 1, 4 will be left in different states corresponding to
different measurement results. For results
$|+^{'}\rangle_{2}|+^{'}\rangle_{3}$,
$|-^{'}\rangle_{2}|-^{'}\rangle_{3}$, ions 1, 4 will be left in
maximally entangled state
$\frac{1}{\sqrt{2}}(|m_{-}\rangle_{1}|m_{+}\rangle_{4}-|m_{+}\rangle_{1}|m_{-}\rangle_{4})$
with probability $\frac{1}{2}|\mu|^{2}|\nu|^{2}|a|^{2}|b|^{2}$.
For the result $|+^{'}\rangle_{2}|-^{'}\rangle_{3}$, ions 1, 4
will be left in the state in Eq.(\ref{stateofrsp}), i.e. the state
Cliff want to prepare remotely. The probability is
$\frac{1}{4}|a|^{2}|b|^{2}(|\mu|^{4}+|\nu|^{4})$. For the results
$|-^{'}\rangle_{2}|+^{'}\rangle_{3}$, ions 1, 4 will be left in
the state
$\frac{1}{2}ab(\mu^{2}|m_{+}\rangle_{1}|m_{-}\rangle_{4}+\nu^{2}|m_{-}\rangle_{1}|m_{+}\rangle_{4})$
with probability $\frac{1}{4}|a|^{2}|b|^{2}(|\mu|^{4}+|\nu|^{4})$.
This state can be converted into the state in
Eq.(\ref{stateofrsp}) by operating $\sigma_{x}$ operations on the
two ions. So if ions 2, 3 are measured in states
$|+^{'}\rangle_{2}|-^{'}\rangle_{3}$,
$|-^{'}\rangle_{2}|+^{'}\rangle_{3}$, the ions 1, 4 will be
prepared in the state in Eq.(\ref{stateofrsp}) successfully with
probability $\frac{1}{2}|a|^{2}|b|^{2}(|\mu|^{4}+|\nu|^{4})$.

Compared with the previous scheme for the remote preparation of
entangled states, the current scheme embeds the following
advantages: (1) it does not need the joint measurement; (2) it can
realize the remote preparation of entangled state for ions by
using linear optical elements. So it is simpler than the cavity
QED or ion-trap schemes.

By far we have only discussed the idea case where we suppose that
a photon impinging on an atom always leads to the process
described in Eq.(\ref{scattering}). But in most case the photon
will not be scattered by the ions, if the ions are placed inside
the MZI. That would mean that detector at the right upper output
port $D_{u}$ will most likely fire. To enhance the scattering
rate, an optical cavity will be added to enclose the MZI. The
detailed description has been discussed in Ref.~\cite{me4}, which
indicates that this cavity will increase the success probability.

Then we will consider the feasibility of the current scheme. As
discussed in Refs. ~\cite{me3, simon}, we can use $^{40}$Ca$^{+}$
as the candidate ion for a possible implementation of the current
scheme. $D_{5/2}$ and $D_{3/2}$ are two metastable levels of
$^{40}$Ca$^{+}$ with lifetimes of the order of $1s$. $s_{1}$ and
$s_{2}$ are two sublevels of $D_{5/2}$ with $m=-5/2$ and $m=-1/2$,
and this two sublevels are coupled to $|e\rangle$ by $\sigma_-$
and $\sigma_+$ light at $854nm$. Here $e, S_{1}, S_{2}, S_{1/2}$
correspond to $e, m_{-}, m_{+}, g$ in Fig.\ref{level}
respectively. That is to say, we use the $S_{1/2}$ as stable
ground state, $S_{1}, S_{2}$ as two degenerate metastable state
and $P_{3/2}$ as excited state. Arbitrary superposition state of
this two degenerate metastable states can be realized by applying
a laser pulse of appropriate length, which can be realized in a
few microsecond~\cite{ion1}. The $^{40}$Ca$^{+}$ in state $S_{1}$
or $S_{2}$ can be excited into the excited state $P_{3/2}$ by
applying one $\sigma_-$ or $\sigma_+$ light at $854nm$. Then decay
from $|e\rangle$ to $S_{1}, S_{2}$, to $D_{3/2}$ and to $S_{1/2}$
are all possible. But Refs.~\cite{ion2, simon} give the transition
probability for $P_{1/2}\rightarrow S_{1/2}(397nm)$ as $1.3 \times
10^8/$s and the branching ratio of $P_{1/2}\rightarrow
D_{3/2}(866nm)$ versus $P_{1/2}\rightarrow S_{1/2}(397nm)$ as
1:15, while the branching ratio for $P_{3/2}\rightarrow
D_{5/2}(854nm)$ versus $P_{3/2}\rightarrow S_{1/2}(393nm)$ can be
estimated as 1:30, giving $0.5 \times 10^7/$s for the transition
probability. So in most case, the $^{40}$Ca$^{+}$ in the excited
state will decay into the stable ground state $S_{1/2}$. The
detection of the internal states of $^{40}$Ca$^{+}$ can be
realized by using a cycling transition between $S_{1/2}$ and
$P_{1/2}(397nm)$~\cite{ion2,ion3}.

To enhance the emission efficiency of the photons from the ions,
we can introduce cavities. Then the following three items will
affect the emission efficiency of the photon from the ions:
\begin{itemize}
    \item The coupling between cavity mode and the $P_{3/2}\rightarrow
    S_{1/2}(393nm)$ transition;
    \item Decay from $P_{3/2}$ to $D_{5/2}$;
    \item Cavity decay.
\end{itemize}
From reference~\cite{decay}, the probability $p_{cav}$ for a
photon to be emitted into the cavity mode after excitation to $e$
can be expressed as $p_{cav}=\frac{4 \gamma
\Omega^2}{(\gamma+\Gamma)(\gamma \Gamma + 4 \Omega^2)}$. where
$\gamma=4\pi c/F_{cav}L$ is the decay rate of the cavity,
$F_{cav}$ its finesse, $L$ its length,
$\Omega=\frac{D}{\hbar}\sqrt{\frac{hc}{2 \epsilon_0 \lambda V}}$
is the coupling constant between the transition and the cavity
mode, $D$ the dipole element, $\lambda$ the wavelength of the
transition, $V$ the mode volume (which can be made as small as
$L^2\lambda/4$ for a confocal cavity with waist
$\sqrt{L\lambda/\pi}$), and $\Gamma$ is the non-cavity related
loss rate~\cite{simon}. From the discussion of Ref.~\cite{simon},
the photon package is about $100ns$, which is a relative long time
for the current scheme to be completed.

When calculating the total efficiency of the current scheme (we
discuss the entanglement concentration via swapping as example),
we must consider the following items:
\begin{itemize}
    \item The emission efficiency of photon: $p_{cav}$, which has included the cavity
    decay; To maximize the $p_{cav}$, we have chosen
    $F_{cav}=19000$, $L=3mm$. Then
    $\gamma=9.9\times10^{6}/s$, $p_{cav}=0.01$~\cite{simon};
    \item The effect of the photon detectors is expressed as
    $\eta$.Here we let a detection efficiency
    $\eta=0.7$, which is a level that can be reached within the current
    technology.
    \item Coupling the photon out of the cavity will introduce
    another error expressed as $\xi$, which can be modulated to be
    close to unit.
\end{itemize}

In addition, we suppose Alice and Bob all have shared an ensemble
of nonmaximally entangled pairs of ions with Cliff. After
considering the above factors, the total success probability can
be expressed as follow:
$P={\frac{a^{2}(1-a^{2})}{2}}\times{p_{cav}}^{2}\times{\eta}\times{\xi}$
for the entanglement concentration via swapping, that is to say,
if we input photon with the rate of $1000000/s$, we can get eight
pairs of concentrated entangled $^{40}$Ca$^{+}$ ions per second
for $a^{2}=0.7$.

In conclusion, we proposed a linear optical scheme for the
teleportation of unknown ionic states, the entanglement
concentration for nonmaximally entangled ionic states via
entanglement swapping and the remote preparation of entangled
states for ions that have never interacted before. The current
scheme does not need the realization of the complex joint
measurement, i.e. the joint measurement has been converted into
the separate measurements on single photons or ions. Quantum
information processing for ions can be realized by using linear
optical elements. In addition, the scheme avoids the complexity of
the ion-trap schemes. However, the current scheme can not realize
the quantum dense coding, because the current scheme can not
discriminate the four Bell states conclusively.

\begin{acknowledgments}
This work is supported by the Natural Science Foundation of the
Education Department of Anhui Province under Grant No: 2004kj005zd
and Anhui Provincial Natural Science Foundation under Grant No:
03042401 and the Talent Foundation of Anhui University.
\end{acknowledgments}

\end{document}